\documentstyle[12pt]{article}

\begin{document}

\baselineskip=7mm

\newcommand{\TeV}{\,{\rm TeV}}
\newcommand{\GeV}{\,{\rm GeV}}
\newcommand{\MeV}{\,{\rm MeV}}
\newcommand{\keV}{\,{\rm keV}}
\newcommand{\eV}{\,{\rm eV}}
\newcommand{\Tr}{{\rm Tr}\!}
\renewcommand{\arraystretch}{1.3}
\newcommand{\be}{\begin{equation}}
\newcommand{\ee}{\end{equation}}
\newcommand{\bea}{\begin{eqnarray}}
\newcommand{\eea}{\end{eqnarray}}
\newcommand{\ba}{\begin{array}}
\newcommand{\ea}{\end{array}}
\newcommand{\bmat}{\left(\ba}
\newcommand{\emat}{\ea\right)}
\newcommand{\refs}[1]{(\ref{#1})}
\newcommand{\ler}{\stackrel{\scriptstyle <}{\scriptstyle\sim}}
\newcommand{\ger}{\stackrel{\scriptstyle >}{\scriptstyle\sim}}
\newcommand{\lag}{\langle}
\newcommand{\rag}{\rangle}
\newcommand{\ns}{\normalsize}
\newcommand{\cm}{{\cal M}}
\newcommand{\gr}{m_{3/2}}
\newcommand{\p}{\partial}

\def\321{$SU(3)\times SU(2)\times U(1)$}
\def\tl{{\tilde{l}}}
\def\tL{{\tilde{L}}}
\def\bd{{\overline{d}}}
\def\tL{{\tilde{L}}}
\def\a{\alpha}
\def\b{\beta}
\def\g{\gamma}
\def\c{\chi}
\def\d{\delta}
\def\D{\Delta}
\def\db{{\overline{\delta}}}
\def\Db{{\overline{\Delta}}}
\def\e{\epsilon}
\def\l{\lambda}
\def\n{\nu}
\def\m{\mu}
\def\nt{{\tilde{\nu}}}
\def\p{\phi}
\def\P{\Phi}
\def\x{\xi}
\def\r{\rho}
\def\s{\sigma}
\def\t{\tau}
\def\th{\theta}
\def\thm{\theta_M}
\def\ne{\nu_e}
\def\nm{\nu_{\mu}}
\def\nt{\nu_{\tau}}

\begin{titlepage}
\title{ {\large\bf NEUTRINOS AND STRUCTURE OF THE INTERMEDIATE MASS SCALE}\\
                                          \vspace{-4cm}
{\ns \it Yukawa Institute, Kyoto}         \hfill{\ns YITP-95-6\\}
                                          \hfill{\ns November 1995\\}
                                          \hfill{\ns hep-ph/9511000\\}
                                          \vspace{3cm} }

\author{
         Alexei Yu.~Smirnov \\[.5cm]
  {\ns\it Yukawa Institute for Theoretical Physics}\\
  {\ns\it Kyoto University, 606-Kyoto, Japan}\\
  {\ns\it International Centre for Theoretical Physics}\\
  {\ns\it P.~O.~Box 586, 34100 Trieste, Italy} \\
  {\ns\it Institute for Nuclear Research, Russian Academy of Sciences}\\
  {\ns\it 117312 Moscow, Russia} }
\date{}
\maketitle
\begin{abstract}
\baselineskip=6.5mm
{\ns
Neutrino data  lead via the see-saw mechanism
to masses of the right handed neutrinos
at the  intermediate mass scale. Simple formalism is suggested
which incorporates the quark - lepton symmetry
and allows one to find
properties of the intermediate scale (masses and mixing)
from  neutrino data.
Averaged mass scale and the mass hierarchy parameter
are introduced and fixed by the data.
They  determine natural ranges
of masses and mixing  at the intermediate scale.
In particular, scenario which includes
the MSW solution of the  solar neutrino  problem
and tau neutrino as  the hot component of
Dark Matter of the Universe leads to
$M_2 = (2 - 4) \cdot 10^{10}$ GeV and
$M_3 = (4 - 8) \cdot 10^{12}$ GeV
in agreement with the linear mass hierarchy. Strong
deviations from the natural ranges imply fine tuning of
parameters or/and certain symmetry of the Majorana mass matrix of
the right handed neutrinos.
}
\end{abstract}
\thispagestyle{empty}
\end{titlepage}

\section{Introduction}

Neutrino data indicate an  existence of the intermediate mass scale
$M \sim 10^{10} - 10^{13} \GeV$.
This statement is based on  the following
assumptions:\\

1. Neutrino masses are generated by the see-saw mechanism \cite{sees}.
Mass matrix of light Majorana neutrinos, $m_{\nu}$, has the
following form
\be
\label{ss}
m_{\nu} \approx - m_D M_R^{-1} m_D^T .
\ee
Here $m_D$ is the neutrino Dirac mass matrix and
$M_R$ is the Majorana mass matrix of the
right handed (RH) neutrino components
\footnote{In models with intermediate scale direct Majorana masses of
left components are typically much smaller than those from (\ref{ss}).}.

2. There is a quark-lepton symmetry (or analogy), according to which
the Dirac mass matrices of leptons are similar to
mass matrices  of quarks. In particular, the  Dirac neutrino
matrix, $m_D$,  is similar to mass matrix of the up quarks:
$
m_D \sim m^{up}
$.
This results in equality or certain  relations between
quark and lepton masses.
The relations  are generic
consequences of
Grand Unification, but they  also often appear in string inspired
models.

3. Deficit of solar neutrinos is due to the resonance flavor conversion
(MSW) $\nu_e \rightarrow \nu_{\mu}$.
Large scale structure of the Universe is explained in terms
of Cold plus Hot Dark Matter (HDM) scenario which implies
at least one (presumably, third) neutrino mass
in the range 2 - 10 eV.
The atmospheric neutrino problem has a solution in
terms of oscillations $\nu_{\mu} - \nu_{\tau}$.
Each of these three hints separately
leads via the see-saw relation,
$M \sim m_D^2/ m$, to the
Majorana masses of the RH neutrinos
in the intermediate scale.\\

What is the origin of the intermediate scale?
Do the  masses of the RH neutrinos
related to  masses
from other sectors of theory at the intermediate scale?

There is a number of proposals:
(i) Intermediate scale can be the
intermediate gauge scale: e.g. the scale of L - R symmetry
violation \cite{ms}.
(ii) It could be related to the Peccei-Quinn symmetry
breaking scale \cite{pq}.
(iii) It was marked that RH neutrino masses are in the
range of the supersymmetry breaking scale
in the hidden sector.
(iv) The intermediate scale can originate from
GUT scale and higher mass scales, e.g. Planck scale,  as
$M_I \sim M^2_{GU}/M_{Pl}$.
(v) On the contrary, it could be constructed from much
smaller scales, when more than two states
from each generation participate in the see-saw.
$M_I \sim M^2_{D}/\mu$
with $\mu \sim m_{3/2}$ and
$M_{D} \sim 10^{7} \GeV$.
(vi) The scale can appear as radiative correction to the GUT scale in
the non-susy GUT \cite{witt}.
(vii) It could be just due to smallness
of the corresponding Yukawa couplings,
although such a possibility seems quite unnatural, since even
for third generation the coupling should be
of order $10^{-3}$.\\

The existence of the intermediate
scale is crucial for  possible  unification
of particles and interactions.
And to identify  the origin of
the scale one needs to know a detailed information on
its structure (masses, mixing).
In the most of studies $M_R$ is fixed by some ansatz or
its structure is related by  symmetry to the
structure of quark mass matrices.
This allows one using the see-saw mechanism to make
predictions of masses of  the
light neutrinos.
In this paper we consider the opposite task:
determination of  masses and mixing of the right handed
neutrinos from the low energy neutrino data.
In the specific context of the  SO(10) model the problem
has been solved in \cite{alb}. In contrast we
will  study  properties  of the intermediate scale within the
general model independent framework (1  - 3).

The paper is organized as follows. In sect.2 the
see-saw mass matrix is defined
and main points of the approach are formulated.
In sect. 3 we get a relation between the
eigenvalues of  mass matrices entering the see-saw formula.
In sect.4  mass parameters in absence
of mixing at the intermediate scale are determined. These
parameters in turn determine  the
average scale and  mass hierarchy.
The possibility of the universal mass
scale is considered in sect.5.
In sect.6 we study the influence of
mixing at the intermediate scale on
 mass hierarchy and  on  mixing of the light neutrinos.
The task is solved for two  (second and third)
generations.
The effect of  first generation is estimated in sect. 7.
Sect. 8 summarizes the results.


\section{The see-saw mass matrix}

Let us  factorize the lepton mixing matrix in the
following way \cite{smi}
\be
\label{lm}
V_l = V_D \cdot V_s \ ,
\ee
where
\be
\label{Dm}
V_D \equiv S_l^{+} \cdot S_{\nu}   \ ,
\ee
and $S_l$, $S_{\nu}$ are the transformations which diagonalize
the Dirac mass
matrices of charge leptons and neutrinos
correspondingly. The matrix
$V_D$ is a direct analogy of the CKM-matrix of quark mixing,
whereas $V_s$ specifies the effect of the see-saw mechanism,
i.e. the effects of the Majorana mass matrix
of the  RH neutrino components. As follows from  (\ref{ss})
$V_s$ is determined by diagonalization
\be
\label{diag}
V_s m_{ss} V_s^+  = diag (m_1, m_2, m_3)
\ee
of matrix \cite{smi}
\be
\label{mss}
m_{ss} = - m_D^{diag} M_R^{-1} m_D^{diag} \ .
\ee
Here $m_i$ (i = 1,2,3)
are the  masses of light neutrinos and
\be
\label{Ddiag}
m_D^{diag} \equiv diag (m_{1D}, m_{2D}, m_{3D})
\ee
is the diagonalized Dirac mass matrix of neutrinos,
$m_{iD}$ are the eigenvalues of $m_D$. In (\ref{mss}) the
$M_R$ is the Majorana mass matrix in the basis
where $m_D$ is diagonal. This basis can be called the Dirac basis.
Further on we will study  properties of the
$M_R$ in this Dirac basis.
For simplicity we will suggest that all  matrices  are
real, although we will admit both negative
and positive values of masses, which corresponds to
different CP-parities of neutrinos.

The Eq. (\ref{mss}) can be considered as
the relation between  mass matrix of light neutrinos,
$m_{ss}$, in the Dirac basis,
 the mass matrix of the RH neutrinos  and
the eigenvalues of the Dirac mass matrix. We will use this relation
to study  properties of the intermediate scale. Indeed,

(1) the eigenvalues of $m_{ss}$ are the masses
of light neutrinos which will be taken from  experiment.

(2) According to the quark-lepton symmetry the
eigenvalues $m_{iD}$ can be related to  masses of upper quarks
\be
\label{ql}
m_{iD} = k_i m_i^{up}
\ee
at some unification scale (e.g. at GU) and $k_i$ are
coefficients of the order  1 - 3.

(3) The mass matrix $m_{ss}$ gives  an additional
contribution to the lepton mixing: $V_s \neq I$. In
two generation case $V_s$ is parametrized by one angle
$\theta_s$ which we  call the see-saw angle
(\cite{smi}) and total lepton mixing is
\be
\label{ws}
\theta_l = \theta_D + \theta_s \ ,
\ee
where $\theta_D$ follows from the Dirac mass matrix
and can be related to the quark mixing angle.
The data on lepton mixing then  restrict
$\theta_s$,  and consequently $M_R$.


\section{Mass Relation}

Majorana mass matrix $M_R$  can be written as
\be
\label{MR}
M_R =  S_R M_R^{diag} S_R^T \ .
\ee
where
\be
\label{MRdiag}
M_R^{diag} \equiv diag (M_1, M_2, M_3) ,
\ee
and $M_i$ (i = 1, 2, 3) are the masses of the RH neutrinos.
For fixed values of $m_i$ and $m_{iD}$ the masses $M_i$ depend
on mixing, i.e. on $S_R$. However,
there is a
relation between masses which does not depend on the mixing.
Calculating the determinants in the LH side and in the RH
side of the see-saw formula (\ref{ss}) one finds
\be
\label{rel}
M_1 \cdot M_2 \cdot M_3 =
\frac{ m_{1D}^2 \cdot m_{2D}^2 \cdot m_{3D}^2 }
{ m_{1} \cdot m_{2} \cdot m_{3} }  \ ,
\ee
where the RH side can
be in principle determined from the experiment.
For example, $m_2$ can be fixed by solar neutrino data,
$m_3$ can be restricted by cosmological data, or by
atmospheric neutrino data or/and  by
accelerator experiments.
However, it will be difficult to get the information
on $m_1$ in the
case of strong mass hierarchy. Moreover,  for $m_1 < 10^{-5}$ eV
the see - saw contribution may  be smaller,  than
e.g. gravitationally induced mass. In this case even known value
$m_1$ is
useless for the determination of the intermediate scale masses.
In this connection we will consider the task for
the second and the third
generations and then estimate possible influence
of the first generation.
The expression (\ref{rel}) can be rewritten as
\be
\label{rel2}
M_2 \cdot M_3 = \xi_1 \frac{ m_{2D}^2 \cdot m_{3D}^2 }
{ m_{2} \cdot m_{3} }  \ ,
\ee
where
\be
\xi_1 \equiv \frac {m_{1D}^2}{M_1 m_1}.
\ee
If $\xi_1 =1$ the task is reduced to two neutrino task.
The influence of the first generation is
strong if $\xi_1$ strongly deviates from 1 (see sect. 7).


\section{Masses at zero mixing. Averaged mass scale and
mass hierarchy parameter}

We will  consider first the case of two generations
suggesting that
$\xi_1$ is close to 1.
Let us introduce two mass parameters
\be
\label{mo}
M_{02}  \equiv  \frac{ m_{2D}^2 }{ m_{2} } , \ \ \
M_{03} \equiv \frac{ m_{3D}^2 } { m_{3} } ,
\ee
where all the masses are taken at the electroweak scale.
Evidently they coincide with  masses of the RH neutrinos in
absence of mixing in $M_R$ (in Dirac basis, where $m_D$ is diagonal).

Masses
$M_{02}$ and  $M_{03}$ can be determined
from low energy data as follows.
Suppose that the solar neutrino problem is solved by small mixing
MSW solution, then in the case of mass hierarchy one has
(see e.g. \cite{krsm})
\be
\label{m2}
m_2 = \left( 2.5 \ ^{+ \ 0.9}_{-\  0.7} \right) \cdot 10^{-3} \eV .\\
\ee
Suppose that
at the Gran Unification scale, $M_{GU} = 2 \cdot 10^{16} \GeV$:
\be
\label{boun2}
m_{2D} = k_2 m_c ,
\ee
where  $m_c$ is the mass of charm quark.
Then from (\ref{mo}) one gets $M_{02}(m_Z)$ at $m_Z$ :
\be
\label{moren}
M_{02}(m_Z)  = \eta_2^2 k_2^2 \frac{ m_{c}^2 (m_Z) }{ m_{2} (m_Z) } ,
\ee
where $\eta_2$ is the renormalization group factor corresponding to
the boundary condition (\ref{boun2}).
At one loop level
\be
\eta_2 \approx \frac{E_{\nu}}{E_c} ,
\ee
where $E_{\nu}$ and $E_{c}$ describe  renormalization effect
respectively for the Dirac neutrino mass and charm mass
between  $m_Z$ and $M_{GU}$. The effect of large
Yukawa couplings from third generation is the same
for $m_{2D}$ and $m_c$,  so that only gauge interactions are important.
Using the MSSM particle content and values $m_c(m_Z) \approx 0.67$ GeV,
$m_2 (m_Z) \approx m_2 (0)$ we get $\eta_2 \approx 0.44$
and
\be
\label{mo2num}
M_{02}  = \left(3.5 \  \pm 1.3 \right) \cdot 10^{10} \ k_2^2  \GeV \ .
\ee
Where the error is rough estimation of  uncertainties in the input
parameters ($\alpha_3$,  etc ).
Since $M_{02}$ does not run up to the intermediate scale
the above value gives $M_{02}$ at $M_{02}$.

Similarly,
$M_{03}$
can be related to the top quark mass:
\be
\label{moren3}
M_{03}  = \eta_3^2 \frac{ m_{t}^2 (m_Z) }{ m_{3} (m_Z) } .
\ee
For third generation of fermions we take $k_3 = 1$, as could
be hinted by $b - \tau$ mass unification. The renormalization
group factor $\eta_3$ is
\be
\eta_3 \approx
\frac{E_{\nu}}{E_c}
\frac{D_{\tau}}{D_b}
\left(\frac{D_{\nu}}{D_t}\right)^3 \ ,
\ee
where $D_i$ describe the Yukawa coupling renormalization effect.
In particular, $D_{\nu}$ is the renormalization
effect of the neutrino Yukawa
coupling between $M_{03}$ and $M_{GU}$. (For this estimation we took
$M_{03} \sim 10^{12}$ GeV). Numerically \cite{fr}:
\be
\label{mo3num}
M_{03}  = \left(2.1 \pm   0.7 \right) 10^{12}
\left(\frac{5 \eV}{m_3}\right) \GeV .
\ee
The cosmological bound on neutrino mass is
$m_3 < 50$ eV (for value of Hubble constant
$h  \sim 0.7$) \cite{ko}
gives according to (\ref{mo3num})
$M_{03} > 2 \cdot 10^{11}$ GeV.
More strong bound can be obtained from
the Large scale structure of the Universe \cite{as}.
Pure hot dark matter scenario contradicts the observed
structure.
The hot dark matter contribution
to energy density should be  at least
two times smaller than the contribution of the could dark matter:
i.e. $\Omega_{\nu} < 0.3$. This gives a conservative
bound on the mass \cite{as}:
\be
m_3 < 15 \eV
\ee
which in turn leads according to  (\ref{mo3num})
to the bound on the Majorana mass
\be
\label{mo3bo}
M_{03}  > 6 \cdot 10^{11} \GeV \ .
\ee
The best fit of Large scale structure of the Universe corresponds to
$m_3 \sim 5$ eV \cite{prim} which results in
\be
\label{mo3bf}
M_{03} = (2.1 \pm 0.7) \cdot 10^{12} \GeV \ .
\ee

For $m_3 \sim 0.1$ eV needed to solve  the
atmospheric neutrino problem in terms of neutrino oscillations
one gets
\be
\label{mo3at}
M_{03} = (1.2 \pm 0.3) \cdot 10^{14} \GeV \ .
\ee \\

In what follows we will consider properties of the
intermediate scale for two scenarios of masses
and mixing of light neutrinos.
(1) ``Solar + HDM" neutrino scenario.  It  incorporates   the
MSW solution of solar neutrino problem
and supplies the
HDM component in the Universe. (2) ``Solar + atmospheric"
neutrino scenario. This scenario  solves simultaneously
the solar and the atmospheric
neutrino problems.
Let us stress that forthcoming  experiments will  allow to
distinguish these two cases.\\

The immediate observation is that at least for $k = 1$
\be
M_{02} \ll M_{03},
\ee
i.e. there is no unique
scale for the RH neutrinos even for ``solar + HDM" scenario. \\

The masses
$M_{02}$ and  $M_{03}$ allow one  to introduce
the average mass scale $M_0$:
\be
\label{av}
M_0 \equiv \sqrt{M_{02} M_{03}} ,
\ee
and the mass hierarchy parameter
\be
\label{hi}
\epsilon_0 \equiv \frac {M_{02}}{M_{03}} \ .
\ee
As we will see these two parameters determine important
characteristics of the intermediate scale:
in particular they
give the natural ranges of masses and mixing in the RH sector.\\

{}From (\ref{mo2num})  and (\ref{mo3bf}) one gets
for scenario ``solar + HDM" at $k = 1$:
\be
\label{par}
M_0 = 2.8 \cdot 10^{11} \GeV  \  ,\ \ \ \
\epsilon_0 = 1.3 \cdot 10^{-2} \ .
\ee
The bound from the Large scale structure of the Universe results in
\be
\label{sc2}
M_0 > 1.3 \cdot 10^{11} \GeV \  , \ \ \
\epsilon_0 <  10^{-1} \ .
\ee
In the case of ``solar + atmospheric" neutrino
scenario one has according to (\ref{mo2num}) and (\ref{mo3at})
\be
\label{sc3}
M_0 = 2.0 \cdot 10^{12} \GeV \  , \ \ \
\epsilon_0 = 2.7 \cdot 10^{-4} \ .
\ee

Due to the relation (\ref{rel}) the average mass scale
determines the product of the masses of the RH neutrinos
for arbitrary mixing in the RH sector:
\be
\label{mr2}
M_2 \cdot M_3 = M^2_0
\ee


\section{Universal scale}

Let us consider first the possibility to have the universal
mass scale for all RH neutrinos
\be
M_{1} = M_{2} = M_{3} \sim 2 \cdot 10^{12} \GeV ,
\ee
where the number is taken from (\ref{mo3bf}).
In this case there is no mixing in the RH sector, $V_s = I$,
and correspondingly, $M_{i} = M_{0i}$, lepton mixing is
determined by Dirac mass matrices.
To get $M_{02} = M_{03}$ in the ``solar + HDM"
neutrino scenario one needs according to (\ref{mo2num}) $k \approx  9$.
The neutrino  mass,
$m_{2D}$, can be enhanced in comparison with quark mass, $m_c$,
e.g. if the element $m^D_{23}$ of  the
neutrino Dirac mass matrix equals
$m^D_{23} \sim   3 m^{up}_{23}$.
This  can be achieved by the contribution of
126-plet in the $SO(10)$ context.

Such a possibility faces however two problems:

(1). Mixing between $\nm$ and $\nt$ turns out to
be strongly enhanced. The contribution from the
neutrino mass matrix is
\be
\theta_{\mu \tau}^{\nu} \sim 3 \sqrt\frac{m_c}{m_t}
\ee
which gives $\sin^2 2\theta_{\mu \tau} \approx 0.14$.
For $m_3 \sim 5$ eV (in the cosmologically
interesting region) this value is already excluded by the
E531 experiment
($\sin^2 2\theta_{\mu \tau} < 5 \cdot 10^{-3}$).
CHORUS and NOMAD will  further strengthen the
bound. To satisfy the bound one should suggest
 strong
cancellation of the contributions from  neutrino and charge lepton
sectors.

(2).  On the contrary,  neutrino contribution to
mixing between the first and the second generations
is suppressed. Even if the element $m_{12}^D$ contains
an additional factor 3 in comparison with corresponding
quark mass,
the contribution to mixing is
\be
\theta_{e \mu }^{\nu} \sim \frac{1}{3} \sqrt\frac {m_u}{m_c} \ .
\ee
The total mixing angle,
\be
\theta_{e \mu } > \sqrt{\frac {m_e}{m_{\mu}}} -
\frac{1}{3} \sqrt{\frac{m_u}{m_c}} \
\ee
is too big: $\sin^2 2\theta_{e\mu} \sim 1.3 \cdot 10^{-2}$
is on the border of  region of small mixing MSW solution
to the solar neutrino problem.

Thus the possibility of the universal intermediate scale
for ``solar + HDM" scenario, although is not excluded,  turns
out to be strongly restricted by present
data and will be checked
by forthcoming experiments.
Universal scale is practically excluded in the
case ``solar + atmospheric" neutrino scenario.


\section{Mass hierarchy and mixing at the intermediate scale}

In the case of two generations the mixing, $S_R$, in the
RH sector is characterized by one angle $\theta_M$.
According to (\ref{mr2}) the product of masses
$M_2 \cdot M_3$ does not depend on $\thm$.
However the masses and the mass  ratio (mass
hierarchy)
\be
\label{eps}
\e \equiv \frac{M_2}{M_3}
\ee
do depend on mixing.
Using the definition (\ref{eps}) and mass relation (\ref{mr2})
we can write
\be
\label{m23}
M_2 = M_0 \sqrt \e , \ \ \
M_3 = M_0 \frac{1}{\sqrt{\e}} \ .
\ee
Mixing in the RH sector changes the splitting between
masses but it does not change the product of masses
(for fixed masses of light neutrinos and Dirac mass terms).

Substituting  (\ref{MR},\ref{mr2}) into  see-saw
formula (\ref{ss}) for two generations
one finds relation between $\thm$ and $\e$
\be
\label{con}
\sin^2 \thm =  \frac {1}{(1 - \e)(1 - \e ^2_D)} \left[
\pm \left( 1 + \frac{m_2}{m_3} \right)
\sqrt{\e _0 \e} - \e - \e ^2_D \right] ,
\ee
where
\be
\e _D \equiv \frac{m_{2D}}{m_{3D}}
\ee
is the mass hierarchy in Dirac sector.
At $m_2/m_3 \ll 1$ and
$\epsilon \gg \epsilon_D^2$ the expression (\ref{con})
reduces to
\be
\label{conn}
\sin^2 \thm \approx \frac {\e}{1 - \e}
\left[
\pm \sqrt{\frac{\e _0}{\e}} -1
\right] \ .
\ee
The latter  does not depend on
$m_2/m_3$ and  $\e_D^2$ explicitly.
These values  enter only via
$\epsilon_0 = \epsilon_D^2/(m_2/m_3)$.\\

Diagonalization of the see-saw mass (\ref{mss}) matrix gives
for the see-saw angle
\be
\label{ssae}
\tan 2 \theta_s =
\frac {\sin 2\thm \e_D (1 - \e)}
{\e - \e _D^2  + \sin^2 \thm (1 - \e)(1 + \e _D^2)}
\ee
Substituting
$\sin ^2 \thm$ from  (\ref{conn})
in this expression we get for
$\epsilon \gg \epsilon_D^2$:
\be
\label{ssa}
\sin^2 \theta_s \approx \frac {\e^2_D}{\e_0} \left[
\pm \sqrt{\frac{\e_0}{\e}} -1 \right] \ .
\ee
The Eq. (\ref{con},\ref{ssae}) or (\ref{conn},\ref{ssa}) are the basic
relations we will use to study the properties
of the intermediate scale.
The effect of mixing in the intermediate scale is different
for $\e > 0$ and $\e < 0$,  i.e. for the cases
of equal and opposite CP - parities of neutrinos.
We will consider these two cases separately.


\subsection{$\e_0 > 0$}

If $\e_0$ is positive then according to (\ref{conn})
$\e$ should be also positive
which means that  neutrinos have the same
CP - parity.
{}From  positivity of the  RH side of Eq. (\ref{conn})
it follows that only sign plus is possible,  and moreover
\be
\e > \e_0 ,
\ee
i.e. mixing at the intermediate scale always enhances the hierarchy
of masses (for fixed values of $m_i$ and $m_{iD}$).

According to (\ref{conn}) mixing angle $\thm$ first
increases with $\e$, then reaches maximum value
\be
\sin^2 2\thm^{max} \approx \e_0 \ \  \rm{at}\ \
\e \approx \frac{\e_0}{4} \ ,
\ee
and then decreases again. It becomes zero at
\be
\label{elim2}
\e \approx  \frac{\e^4_D}{\e_0} \ .
\ee
For very strong mass hierarchy,
$\e \ll \e_D^2$,  the expression for  mixing angle
can be approximated as
\be
\label{verysm}
\sin^2 \thm \approx  \sqrt{\e _0 \e} - \e - \e_D^2 .
\ee
Thus  there is maximal value of mixing,
$\thm \approx  \sqrt{\e_0} /2$ ,
in the RH sector
determined by parameter $\e_0$.
The hierarchy parameter is restricted by
$\e^4_D/\e_0 \leq \e \leq \e _0$.

The values of angle and mass hierarchy
\be
\thm \sim   \frac{\sqrt{\e_0}}{2} \ , \ \ \
\e \sim \left( \frac{1}{4} - \frac{1}{2} \right) \e_0
\ee
can be considered as the natural values.
They do not imply  any fine tuning of  parameters
and any additional symmetry in $M_R$.

In the limit of very strong hierarchy, $\e \ll \e_0$
one has from (\ref{verysm})
\be
\label{lim}
\sin^2 \thm \approx
\sqrt{\e_0 \e}  \gg \e .
\ee
This inequality implies smallness of
the determinant of $M_2$:
\be
\label{det}
\frac{Det M_R}{M_{23}^2} \approx \frac{\e}{\sin^2 \thm}
\sim \sqrt{\frac{\e}{\e_0}} \ll 1 \ ,
\ee
i.e. near to singular character of $M_R$.
This in turn means fine tuning
of  elements of the matrix in the Dirac basis.\\

Let us consider now the dependence of the see-saw angle
on the mass hierarchy.
According to (\ref{ssa}) the angle is zero at $\e = \e_0$,  it
increases with diminishing $\e_M$.
In the range
$4\e_D^4/\e_0   \ll  \e \ll \e_0$ the dependence can be approximated
by (see (\ref{ssa})):
\be
\label{ssalim}
\sin^2 \theta_s \approx \frac {4 \e^2_D}
{\sqrt{\e_0 \e}}
\ee
(this formula is true when
$\sin^2 2\theta_s$
is still
much smaller than 1).
With further decrease of $\e$ the mixing  approaches maximal value
($\sin^2 2\theta_s = 1$) at
\be
\label{maxa}
\e =  \frac{4 \e^4_D}{\e_0},
\ee
and then decreases up to zero at
$\e =  \frac{\e^4_D}{\e_0}$   (see (\ref{verysm}) ).
Very strong hierarchy leads  to strong
see-saw enhancement of lepton mixing \cite{smi}. \\

Let us consider the applications of the results.
The  natural value of mass hierarchy
in the case ``solar + HDM" neutrino scenario is
$\epsilon \sim \epsilon_0/4 \sim (3 - 4) \cdot 10^{-3}$.
It coincides with hierarchy in Dirac sector:
$
\e \approx \e_D
$.
Thus ``solar +  HDM" neutrino
scenario implies  linear hierarchy
of masses of the RH neutrinos.
This may be considered as a hint to the common origin
of the Yukawa couplings which generate  the
Dirac and the Majorana mass matrices.
Numerically, for $\e = \e_0/3$ we get
$M_2 = 2 \cdot 10^{10}$ GeV
and  $M_3 = 5 \cdot 10^{12}$ GeV . These masses could be
generated by the interaction with scalar field acquiring
the VEV $V \sim 10^{13}$ GeV. \\

Lower bound on mass hierarchy follows from experimental bound on
mixing between the second and the third generations.
Indeed, if there is no strong cancellation
between the Dirac matrix and see-saw matrix contributions, so that
\be
\sin^2 2\theta_s <
\sin^2 2\theta_{\mu \tau} ,
\ee
then one gets according to (\ref{ssa})
\be
\label{elim3}
\e > \frac{\e_0}
{
\left(1 +
\sin^2 2\theta_{\mu \tau} \cdot
\frac{\e_0}{4\e^2_D}\right)^2
}
\ee
and present experimental bound:  $\sin^2 2\theta_{\mu \tau}$ at
$\Delta m^2 > 25 \eV ^2$ gives
\be
\e > \frac{\e_0}{9} .
\ee
For $k = 1$ this leads to  bounds on the RH masses
$M_2 > 10^{10} \GeV$ and
$M_3 < 10^{13} \GeV$.
CHORUS and NOMAD will further
strengthen the bound on hierarchy up to $\e > (1/4 - 1/2)\e_0$
and therefore squeeze the intervals for masses.\\

In principle, the mass $m_3$  in the cosmologically interesting region
does not necessarily imply $M_{33} < 10^{13}$ GeV .
Admitting the tuning of the $Det M_R$  (\ref{det}) and
strong cancellation
between the the Dirac and the see-saw contributions to the
lepton mixing (to satisfy the experimental bounds)
one can enhance the hierarchy
and therefore push $M_{3}$ to higher values.
The increase of $M_{3}$
diminishes (removes) the renormalization effect
due to large Yukawa coupling of
neutrino from the third generation \cite{fr,mur}.
According to  (\ref{m23}) the mass $M_3 \sim 10^{16}$ GeV can
be achieved for the hierarchy parameter
$\e = (M_0/M_{GU})^2 \sim 10^{-8}$. In this case
as follows from (\ref{det})
the level of fine tuning is $10^{-3}$. \\

On the contrary, in ``solar + atmospheric" scenario
one can use the limit
of strong mass hierarchy to enhance the mixing and thus
to explain the atmospheric neutrino deficit.
According to (\ref{maxa})
for $\e_D = 3\cdot 10^{-3}$ and $\e_0 = 2.7 \cdot 10^{-4}$
maximal see-saw mixing is achieved at
$\e \sim 10^{-6}$. This corresponds to the
mass of third neutrino $M_3 \approx 2 \cdot 10^{15} \GeV$ ,
i.e. of the
order of $M_{GU}$. Second neutrino becomes rather light:
$M_2 \sim 2 \cdot 10^{9} \GeV$. There is no usual intermediate scale
in this case.
In this connection let us mark the following possibility.
At the Grand Unification scale only $M_3$ acquires  mass,
whereas
$M_2$ and $M_1$ are massless. Masses of the
second and first neutrinos appear as the result of violation
of certain horizontal symmetry,
by e.g. nonrenormalizable interactions, so that $M_2 \sim M_{GU}
(M_{GU}/M_{Pl})^2$.


\subsection{$\e < 0$}

According to (\ref{conn})
$\e_0$ should be  also negative.
However, as follows from (\ref{conn})
the mass hierarchy can be both stronger (sign plus in front of
square root) and weaker (sign plus and minus)  than
$\e _0$. Let us consider the case of weaker
hierarchy:
\be
|\e| > |\e_0|
\ee
There is no  upper bound on mixing angle $\thm$.
With increase of $\e$ the $\sin^2 \thm$ increases; it equals
\be
\sin^2 \thm = \frac{1}{2} (1 \pm \sqrt{|\e_0|})
\approx \frac{1}{2}
\ee
at  $|\e| = 1$, i.e. for equal absolute values of the
RH neutrino masses. Thus to have $|M_2| \approx |M_3|$
one needs practically maximal mixing in the RH mass matrix.
Note that in all region $\sin^2 \thm \sim \e$ , i.e.
naturalness criteria is fulfilled. \\

The see-saw angle is determined by
\be
\label{ssa2}
\sin^2 \theta_s \approx
\frac{\e^2_D}{\e_0} \left[ 1 - |\e_0|
- \sqrt{\frac{\e_0}{\e}}(1 - |\e|) \right]
\approx \frac {\e^2_D}{\e_0}  \approx \frac{m_2}{m_3}
\ee
for $|\e| \sim 1$. Using the value $\e_0$ (\ref{par}) we
find $\sin^2 2\theta_s \approx 3 \cdot 10^{-3}$
which is near the
existing experimental bound for large mass splitting
(``solar + HDM").
In the case $|\e| \sim 1$ the mass matrix $M_R$ is strongly
off diagonal:
$M_{22} = - M_{33} \propto
\sqrt{\e_0}$
and
$M_{23} \propto \sqrt{1 - \e_0}$.

Mixing allows one to relax mass hierarchy
and therefore diminish the
value of $M_3$. In this connection let us mark two examples in the
``solar + atmospheric" neutrino scenario.

(1). Linear mass hierarchy:
$\e \sim 3 \cdot 10^{-3}$. In this case for $k = 1$
we get
$M_3 = 3\cdot 10^{13} \GeV$,
$M_2 = 1.2 \cdot 10^{11} \GeV$,
$\sin^2 \thm = 2 \cdot 10^{-3}$, and
$\sin^2 2\theta_s = 2 \cdot 10^{-3}$.

(2). Equal masses:
$|M_2|  \sim |M_3| \sim M_0$. In this case diagonal elements of
matrix $M_R$  are suppressed in comparison with off diagonal by
$\sqrt{\e_0} \approx 10^{-2}$. The largest element of the mass matrix
$M_{23} \approx M_0 \approx 2 \cdot 10^{12} \GeV$.
If $m_2 = 4 \cdot 10^{-3} \eV$
and  $m_3 = 3 \cdot 10^{-2} \eV$ then according to (\ref{ssa2})
$\sin^2 2\theta_s \sim 0.5$
in the region of solution of the atmospheric neutrino problem.

Thus it is possible to explain simultaneously both the solar and the
atmospheric neutrino problems by pseudo Dirac structure
at the intermediate scale
$M = (2 - 3) \cdot 10^{12} \GeV$.
Such a structure can be obtained by  imposing, e.g., $U(1)_G$
horizontal symmetry
with charge prescription  $ (0, -1, +1)$ , then scalar with
zero G- charge will produce the mass matrix
\be \label{mm1}
 M_R = \bmat{ccc} M_1 & 0 & 0\\
                         0 & 0 & M \\
                         0 & M & 0
              \emat \;.
\ee
The violation of this symmetry  should be
characterized by factor $10^{-2}$.

\section{Effect of first generation}

If mixing of the first generation
with second and third is sufficiently small, then the
task is reduced to two
neutrino task
and the results for three generations  coincide
with those obtained in the previous sections.
Smallness of mixing needed to solve
the solar neutrino problem by the MSW effect
may indicate on such a weak influence.
Let us consider
situations  when  influence of the first
generation is strong.

As we marked in sect.3 the influence of
first generation can be characterized by parameter
$\xi_1$  in such a way  that strong deviation of
$\xi_1$ from 1 means  the strong effect of the first generation.

Simple dependence of
$\xi_1$ on the matrix element $M_{12}$
can be found explicitly for the case
when first generation mixes with second generation only.

Parameter
$\xi_1$ as  function of $M_{12}$
is different for two cases depending on whether
$M_{22}$ is larger or smaller than $M_{11} / \e^{'2}_D$,
where $\e^{'2}_D \equiv m_u/m_c$. \\

\noindent
(a). $M_{22} < M_{11} / \e^{'2}_D$.
Main features of the dependence $\xi_1(M_{12})$
are the following.\\
(i)  $\xi_1 = 1$ at $M_{12} = 0$ ( the first family decouples). \\
(ii) $\xi_1$ increases with $M_{12}$ and its dependence
can be approximated by
\be
\label{xi2}
\xi_1 \sim
\frac{M_{11} M_{22}}{ D_{12}},
\ee
where $D_{12} \equiv Det M \equiv M_{11} M_{22} - M_{12}^2$
in the region
\be
M_{12} \sim \sqrt{M_{11} M_{22}} .
\ee
According to (\ref{xi2})
$\xi_1 \rightarrow \infty$, when $D_{12} \rightarrow 0$.\\
(iii) At  $M_{12} > \sqrt{M_{11} M_{22}}$,  the parameter
$\xi_1$ changes the sign and its  absolute value decreases with
further increase of $M_{12}$. For example at
$M_{12} \sim M_{22}/2$:
\be
\xi_1 = \frac{\e_D^2 M_{22}}{M_{11}}
\frac{1}{(\sqrt2 - 1)^2}
\ee
and if $M_{11}/M_{22} \sim \e_D$ one gets
$\xi_1 \ll 1$ . \\
(iv)  For $M_{12} \gg M_{22}$, i.e.
when nondiagonal element dominates one gets:
$
\xi_1 \rightarrow \e_D .
$
\\

\noindent
(b).  $M_{22} > \frac{M_{11}}{\e^{'2}_D}$.  (This case covers
in particular the mass matrices
satisfying Fritzsch ansatz ($M_{11} = 0$)).
Now $\xi_1$ monotonously decreases from
$\xi_1 = 1$  at $M_{12} = 0$
to $\xi_1 \approx \e_D$ for
$M_{12} \gg M_{22}$. In particular, for
$M_{12} \sim \sqrt{M_{11} M_{22}}$
the
$\xi_1$ dependence can be approximated by
\be
\xi_1 \approx
\frac{M_{22} + \frac{1}{\e_D^2} M_{11}}
{M_{22} + M_{11}}
\ee
and evidently $\xi_1 < 1$ in contrast with the case (a).

%
%

Thus $\xi_1$ strongly deviates from 1
and there is a strong
influence of the first generation  in two cases:

(1)
If $D_{12} \rightarrow 0$ then
$\xi_1 \rightarrow \infty $.

(2)
If $M_{12} > M_{22}$ then
$\xi_1 \ll 1$.

The bound on the mixing of light neutrinos restricts the effect
of the first generation.
Diagonalizing  $m_{s}$ explicitly, we get
the see-saw mixing between first and second generations
\be
\label{tan}
\tan 2 \theta_s^{12} =
\frac{2 \e'_D M_{12}}
{M_{11} - \e_D^{'2} M_{22}}.
\ee
Solution of the solar neutrino problem implies
$m_1 < m_2$ and therefore
$\e^{'2}_D M_{22} < M_{11}$. That is only the case (a)
gives the solution of the problem.
For small mixing solution the angle
$\theta_s^{12}$  can be restricted, if the Dirac mixing
is similar to that
in quark sector. In this case  one has
$\tan 2 \theta_s^{12} < \tan 2\theta_{\odot} \approx 0.1$
(the value needed to explain the
solar neutrino problem) and then from (\ref{tan})
\be
\label{mixbo}
\frac{2 \e'_D M_{12}}{M_{11}} < \tan 2 \theta_{\odot} \sim 0.1.
\ee \\

Let us consider the applications of the results to our
analysis in sect. 2 - 6.

(1) If $D_{12} \rightarrow 0$,  and therefore
$\xi_1 \rightarrow \infty$,  the average scale increases:
\be
M_2 \cdot M_3 = \xi_1 M_0^2 .
\ee
Substituting
$M_{12} \sim \sqrt{M_{11} M_{22}}$
into  (\ref{mixbo}) we find
\be
\label{ineq}
\frac{M_{11}}{M_{22}} > \frac{4 \e _D^{'2}}{\tan^2 2\theta_{\odot}}
\sim 5 \cdot 10^{-3},
\ee
i.e. $M_{11} \gg \e^{'2}_D M_{22}$ which gives
$m_2  \approx m_c^2/ M_{11}$. From this relation we have
\be
M_{22} = M_{02} \left(\frac{M_{11} M_{22}}{D_{12}}\right) .
\ee
Thus diminishing $D_{12}$ one can push the value of $M_{22}$
and therefore $M_2$ up. In this case also $M_{11}$ should increase
according to (\ref{ineq}). The mass $m_1 \approx m_u^2/M_{11}$
turns out to be small. Numerically one may have
$M_{22} \sim M_{33} \sim  2 \cdot 10^{12} \GeV$ ,
$M_{11} > 10^{10} \GeV$ and
$M_{12} > 1.4\cdot 10^{11} \GeV$ .

(2) If $M_{12} > M_{22}$ then
$\xi_1 \ll 1$ which leads to  decrease of
$M_2 \cdot M_3$. Bound on the see-saw mixing angle results in
restriction $M_{12} /M_{11} < \tan 2\theta_{\odot} /(2 \e_D) \sim 16$
and this leads to bound on $\xi_1$.

\section{Conclusion}

1. Simple relations have been derived
between parameters of the Intermediate scale and
neutrino data in  context of
the see-saw mechanism of the
neutrino mass generation and quark - lepton symmetry.\\

\noindent
2. Neutrino masses hinted  by the solar neutrino data,
the Large Scale Structure of the Universe or atmospheric
neutrino anomaly
allow to introduce mass parameters
$M_{02}$ and   $M_{03}$ which in turn determine the
average mass scale
$M_0$
and the mass  hierarchy
$\e _0$.

The scale $M_0$ and  the mass hierarchy
$\e _0$ fix natural ranges of masses
$M_2, M_3$ and mixing of the
RH neutrinos.
Strong deviations from these
natural ranges imply certain symmetry or/and
fine tuning of  elements of $M_R$.
Namely,  the mass matrix should be strongly off diagonal,
which may be stipulated by  certain horizontal symmetry,
or its determinant should be much smaller than
nondiagonal element squared (in the basis where
neutrino Dirac mass matrix is diagonal). This implies
certain correlation between Yukawa coupling generating
the Dirac and the Majorana mass matrices.\\

\noindent
3. Using the solar, atmospheric, and cosmological data
one can make the following tentative conclusions:\\

(a). In the case of  ``solar + HDM" scenario
with $\e > 0$ the masses of the RH neutrinos are restricted by
the following intervals:
$M_2 = (1 - 4) \cdot 10^{10} \GeV$ and
$M_3 = (2 - 8) \cdot 10^{12} \GeV$.
The mass hierarchy is bounded by
$\e = (1/9 - 1) \e _0$, where  the upper edge  corresponds
to the absence of mixing in the RH sector, and the lower edge follows
from
the experimental bound on mixing of light neutrinos.
Thus one gets the
linear hierarchy of RH neutrinos masses:
$M_i \propto m_{iD} \sim m_{i up}$ with
largest mass
$M_3 \approx 5 \cdot 10^{12} \GeV$.
This may testify for (1) simple relation (equality)
of the Yukawa couplings which generate  masses
 of up quarks  and
Majorana masses of RH neutrinos, for  (2) spontaneous violation
of symmetry of lepton number at $V \sim 10^{13} \GeV$.

Unique intermediate scale is not excluded for this scenario,
but it implies rather strong deviation from simple
relations between quark and lepton masses
and it is on the border of already
existing  bounds.

(b). ``Solar + atmospheric" neutrino scenario implies in general
rather strong mass hierarchy in the RH sector. However
if $M_R$ has essentially off diagonal structure, all
the masses of the intermediate scale below
$10^{13} \GeV$ are not excluded.

Another possibility  is that  the
largest mass $M_3$ is of the order of $M_{GU}$. Then
second mass
should be below $3 \cdot 10^{9} \GeV$. It can be generated by
high order nonrenormalizable interactions.
In this case there is strong see-saw
enhancement of lepton mixing which
allows one to solve the atmospheric  neutrino problem.

\noindent
4. First generation  can
strongly influence the average scale
for second and third generation  in two cases:
(i) when $2\times2$ matrix  including first family
is strongly off diagonal
(ii) when the determinant of this matrix is much smaller that the
off diagonal matrix element squared.
In the former case the average scale decreases in the
latter - increases.
In the most natural case
(taking into account small mixing solution
of the solar neutrino problem)
the influence of
the first generation is not strong
and results for two heavy generations can be changed by
factor of the order 1.


\end{document}